# Detecting T Tauri disks with optical long-baseline interferometry [1]

F. Malbet[†,‡] & C. Bertout[†]

[†] *Laboratoire d'Astrophysique, Observatoire de Grenoble*
BP 53, 38041 Grenoble cedex 9, France.

[‡] *Jet Propulsion Laboratory, California Institute of Technology*
MS 306-388, 4800 Oak Grove Drive, Pasadena, CA 91109, U.S.A.

## Abstract

We present synthetic images of disks around T Tauri stars. We calculate visibility curves in order to study the possibility of directly detecting thermal emission from T Tauri disks. Total fluxes of T Tauri disks are compared to the sensitivity of *VISA*, the interferometric sub-array of the European *Very Large Telescope*. We conclude that thermal emission from circumstellar disk around T Tauri stars is detectable with current or soon-to-be interferometric techniques at wavelengths longer than 2.2 $\mu$m. *Subject headings:* accretion disks - techniques: interferometric – stars: circumstellar matter – stars: imaging – stars: pre-main sequence – infrared: stars

## 1. Introduction

T Tauri stars (hereafter TTSs) are low-mass stars in an early stage of stellar evolution. They are believed to originate from the collapse of rotating molecular clouds, which results in the formation of a protostellar core and a flat axisymmetric envelope. The protostellar core becomes a TTS and the envelope forms an optically thick circumstellar disk surrounded by a larger optically thin envelope. Circumstellar disks are supposed to be the reservoir for the formation of planetoids which later evolve in planets. Detecting T Tauri disks is therefore of importance for understanding the formation and the frequency of planetary systems.

As discussed in the next section, numerous indirect arguments support the presence of circumstellar disks around TTSs. There are two major processes of light emission in disks: thermal emission and stellar light scattered in the disk. The first process depends on the radial temperature distribution (originating from accretion energy and from stellar light absorption) and the second one depends on the geometry of the upper optically thin part of the disk atmosphere. In the case of optically thick thermal emission, the shape of the radial distribution of light depends on the black-body emission regime. As discussed in Section 4., the thermal emission drops dramatically in the outer part of the disk (Wien regime) and therefore the thermal emission is only detectable in the inner part of the disk or at long wavelengths. The intensity of scattered radiation varies radially as a power-law and becomes dominant when the thermal emission is in the Wien regime. Scattered radiation is emitted over a large region so that its flux is diluted and can be neglected in parts of the disk where thermal emission is dominant. Moreover, since scattering varies as $\lambda^{-2}$, the scattered emission becomes fainter in the near-infrared range. Scattered emission can be studied at scale of 0.1–10″ with high angular resolution techniques like adaptive optics and speckle interferometry but is almost absent at scale of 0.1–10 mas for very high angular resolution techniques such as optical long-baseline interferometry. Since we are interested here mainly in predictions of interferometric detectability of TTS disks, we only consider the thermal emission of disks.

Long baseline optical interferometers are made of 2 or more telescopes. Interferometers measure the complex visibility $V_i$ corresponding to the spatial frequencies $\vec{B}_i \times \vec{s}/\lambda$, where $\vec{B}_i$ are the baselines of the interferometer and $\vec{s}$ is the normal vector pointing to the observed object ($\vec{B}_i \times \vec{s}$ is the baseline projected onto the sky plane). The maximum resolution for a baseline of length $B$ is $B/\lambda$. The complex visibility is

---

simply the Fourier transform of the object intensity distribution. If the $(u, v)$ plane coverage is sufficient ($u$ and $v$ are the spatial frequencies associated by Fourier transform with the angular coordinates), it is then possible to invert the complex visibility and to reconstruct the image of the object. Although interferometric imaging is carried out commonly with radio interferometers with many antennas, this is not yet possible with an optical or near-infrared interferometer. For wavelengths less than 10 $\mu$m, one could use heterodyne interferometric techniques as in radio-astronomy. However narrow spectral channels are necessary to mix the signal with the local oscillator and the sensivity of the interferometer becomes very low. For instance, the *Infrared Spatial Interferometer* (Townes 1984; Bester et al. 1994) is limited to the 20-30 brightest stars. This is the reason why direct combination of the stellar beams is preferred at wavelengths smaller than 10 $\mu$m. The main limitation is then due to atmospheric fluctuations. The atmosphere becomes more and more perturbative as $\lambda$ decreases (Roddier 1979), and one must then go to short exposures ($t \leq 10$ ms) and small apertures ($D \leq 0.5$ m) and/or active optics (adaptive optics and fringe-tracker) with larger aperture ($D \approx 2$ m). As a consequence, the $(u, v)$ plane coverage is very sparse and image reconstruction is not always possible, at least in the first stages of the construction. In less than a decade, optical interferometric imaging is likely to be developed (e.g. the European *Very Large Telescope Interferometer* and the American *Keck Interferometer*), but the first interferometric data will be limited to visibility measurements. The purpose of this paper is to show that T Tauri disks are detectable even with only the visibility measurements of currently planned interferometers.

We present synthetic images of thermal emission of T Tauri disks that allow us to compute interferometric visibilities. We show that thermal emission of T Tauri disks can be detected by infrared interferometry with baselines in the range of $\approx 100$ m. Section 2. presents current indirect evidence for the presence of T Tauri disks. Section 3. gives details on disk simulations and image computation and presents the interferometric visibilities obtained from the images. Section 4. discusses the possibility of disk detection.

## 2. Indirect evidence for accretion disks around young stars

The idea that TTSs are surrounded by, and interact with, accretion disks has gained large momentum since 1988. The main advantages of the accretion disk idea are

- to explain a variety of apparently unrelated phenomena (UV excess, IR excess, visual extinction, veiling of optical spectrum, Balmer jump, forbidden line profiles) with a single physical process;

- to provide TTSs with a large reservoir of potential energy for covering their radiative expenditures and for driving the powerful T Tauri wind. There is a clear correlation between the forbidden line flux (a mass-loss rate indicator) and the infrared excess of TTSs, which is a measure of mass-accretion via a disk (Cabrit et al. 1990); this connection between accretion and wind is a topic of current work (e.g., Shu et al 1994; Ferreira and Pelletier 1994).

There are two main lines of indirect evidence for the presence of circumstellar disks in TTSs. The first one is the forbidden line profile shape, often blue-shifted with little or no emission on the red line side. Since forbidden emission probes the outer parts of the ionized TTS wind, the observed lack of red-shifted emission shows that the part of the wind which goes away from us must be heavily obscured ($A_v \geq 6$ mag). That is, there is a disk-like screen between us and the receding part of the wind. This result, independent of any assumed wind geometry, is based only on the assumption that the forbidden lines are broadened by Doppler shifts due to the wind, a natural hypothesis since the line widths are several hundred km/s (Appenzeller et al. 1984; Edwards et al. 1987).

An estimate of the occulting disk size can be found from the observed [SII] line flux; the radius of the emission region is typically 10 - 100 AU. With the additional assumption that the outflowing gas has reached terminal velocity in the forbidden line emission region, one finds mass-loss rates in the range $10^{-9}$ - $10^{-7}$ $M_\odot$/yr, comparable to rates derived from other diagnostics, notably H$\alpha$. The forbidden line argument is perhaps the most convincing evidence for the presence of disks because of its very simplicity.

Models of TTS spectral energy distributions bring another line of evidence that cannot be easily dismissed, although it is possible to argue that the theoretical models are too simplistic. The idealized model of a T Tauri system consists of (i) a central, late-type star surrounded by (ii) a geometrically thin, dusty accretion disk that interacts with the star via (iii) a boundary layer (cf. Bertout, Basri & Bouvier 1988).

The best measure of accretion rates in T Tauri disks comes from the blue and ultraviolet spectral ranges. The Balmer continuum emission jumps observed in TTSs can easily be explained if the boundary layer is optically thin in the Paschen continuum (Basri & Bertout 1989). Even more important, the amount of energy available depends only on the accretion rate and not on the star's resources, while the disk paradigm explains naturally the observed correlation between the respective amounts of infrared and ultraviolet excesses. Current estimates for mass-accretion rates in the disk range from a few $10^{-9}$ to a few $10^{-7}$ $M_\odot$/yr (Bouvier & Bertout 1992), i.e., they are comparable to mass-loss rates in TTS winds.

While observed infrared spectral energy distributions (SEDs) of some TTSs are consistent with those of the classical accretion disk model, most are actually flatter: the average slope of TTSs infrared spectra beyond 5$\mu$m or so is proportional to $\lambda^{-3/4}$ rather than to $\lambda^{-4/3}$ as predicted by the standard accretion disk model (Rydgren & Zak 1987). This discrepancy led Adams, Lada, & Shu (1988) to hypothesize that the temperature distribution in T Tauri disks is flatter than in standard accretion disks. Following this suggestion, it has become common practice to parameterize the temperature law index and to use different values of this parameter to model infrared distributions of young stellar objects, particularly those with flat infrared spectra (e.g., Adams, Emerson & Fuller 1990; Beckwith et al. 1990). The radial temperature law is then given by $T \propto r^{-q}$, with $q$ in the range 1/2 (flat infrared SED) to 3/4 ("classical" accretion disk infrared SED). With this approach and an assumed disk radius of 100 AU, one derives typical disk masses of a few percent of one solar mass, with considerable uncertainty.

## 3. Synthetic thermal images and visibilities of circumstellar disks

Work presented in this paper is part of a wider effort to obtain synthetic images of T Tauri disks in order to compare them with future observations (Bertout & Bouvier 1988, Malbet et al. 1992, Bouvier et al. 1992, Monin et al. 1993). Earlier studies concentrated on vertical structure and scattering processes in T Tauri disk. Here we focus our attention on the output images and on their transformation to interferometric visibilities.

A first disk model used for computing images is based on the work of Bertout, Basri & Bouvier (1988). The disk is supposed to be flat and has a radial structure derived from Lynden-Bell & Pringle (1974) with temperature law proportional to $r^{-3/4}$. We did not take into account the flux coming from the interaction region between disk and star, since it emits mainly in the ultraviolet and is located very close to the surface of the star. The boundary layer is not resolved except perhaps in the ultraviolet range which is beyond current interferometry. The flux from the boundary layer is negligible at visible and infrared wavelengths. The effect of shadowing of the inner part of the disk by the star has been taken into account for the central pixel. The disk is supposed to be optically thick even in the outer part of the disk. The star is an unresolved black body.

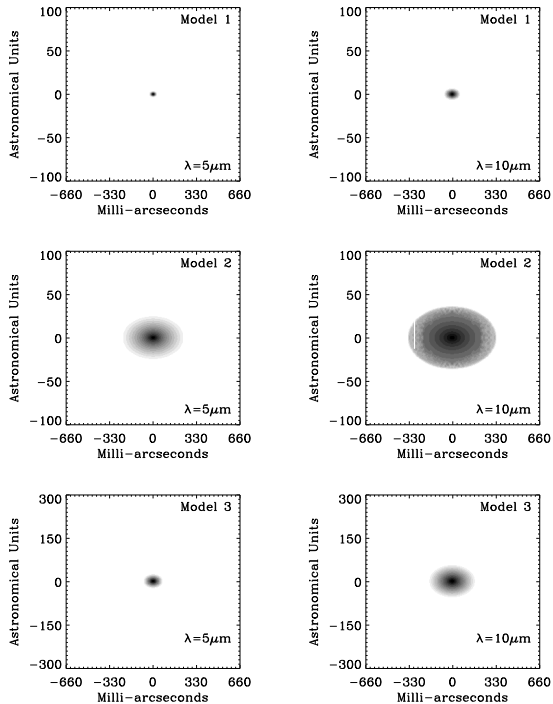

Fig. 1.— Synthetic thermal images of circumstellar disks at $\lambda = 5$ $\mu$m and $\lambda = 10$ $\mu$m (logarithmic scale). Model 1 (classical T Tauri star) is displayed in top panels, model 2 (flat infrared spectrum T Tauri star) in middle panels and model 3 (FU Orionis star) in bottom panels. Left panels are images at 5 $\mu$m, and right panels are images at 10 $\mu$m.

A second model used to compute images assumes the disk has a temperature radial distribution proportional to $r^{-1/2}$. The total flux from the disk is parameterized by the accretion rate which is is needed to get the same total flux with a classical disk model.

Most of the difficulty involved in building an image comes from the linear sampling on the sky. The original model sampling is polar, linear in angle from the star, and logarithmic in radius. We have to transform this in a square linear sampling on the sky without losing or creating any flux. We do this by building a linear regular grid on the sky with a cell size such that several points from the original grid are found inside this cell. Then the solid angle of the original cells is computed in order to conserve the flux. The computation leads to a data cube with two spatial dimensions and a spectral dimension. The spectral energy distribution is calculated with the original sampling and with the regular grid, and we check that the two spectral distributions are identical. The star flux contribution is added in the central pixel.

We simulated 3 different models of circumstellar disks around the same star. The star has radius 3 $R_\odot$, mass 1 $M_\odot$ and 4000 K effective temperature. A 45° disk inclination is assumed.

- Model 1 is a "classical" accretion disk with $10^{-7}$ $M_\odot$/yr accretion rate located in the Taurus-Auriga cloud at 150 pc.

- Model 2 is a disk with a flat infrared spectrum located at the same distance as Model 1.

- Model 3 is a FU Orionis disk located in the Orion molecular cloud located at 450 pc with a $10^{-4}$ $M_\odot$/yr accretion rate.

Disk parameters used in these models are summarized in Table 1. Images obtained at 5 $\mu$m and 10 $\mu$m are displayed on Fig. 1. The $x$-axis gives the spatial scale on the sky in milli-arcseconds and the $y$-axis the absolute spatial scale in Astronomical Units. The emergent spectral energy distributions are presented for each model respectively in the top left panel of Figs. 2, 3 and 4.

Table 1: Disk model parameters.

|  | Model 1 | Model 2 | Model 3 |
|---|---|---|---|
| Inner Radius | 1.1 $R_*$ | 1.1 $R_*$ | 1.1 $R_*$ |
| Outer Radius | 50 AU | 50 AU | 150 AU |
| Distance | 150 pc | 150 pc | 450 pc |
| Temperature | $\propto r^{-3/4}$ | $\propto r^{-1/2}$ | $\propto r^{-3/4}$ |
| Accretion rate | $10^{-7}$ $M_\odot$/yr | $10^{-6}$ $M_\odot$/yr | $10^{-4}$ $M_\odot$/yr |

Table 2: Visibilities for 50m and 100m baselines.

|  | Model 1 | | Model 2 | | Model 3 | |
|---|---|---|---|---|---|---|
| Baseline | 50m | 100m | 50m | 100m | 50m | 100m |
| $\lambda = 0.5$ $\mu$m | 1.0 | 1.0 | 1.00 | 0.99 | 0.96 | 0.87 |
| $\lambda = 1$ $\mu$m | 1.0 | 1.0 | 0.99 | 0.98 | 0.96 | 0.87 |
| $\lambda = 2.2$ $\mu$m | 0.99 | 0.98 | 0.96 | 0.88 | 0.94 | 0.81 |
| $\lambda = 5$ $\mu$m | 0.99 | 0.96 | 0.86 | 0.77 | 0.92 | 0.74 |
| $\lambda = 10$ $\mu$m | 0.99 | 0.95 | 0.66 | 0.40 | 0.87 | 0.65 |
| $\lambda = 20$ $\mu$m | 0.97 | 0.90 | 0.41 | 0.21 | 0.82 | 0.56 |

In order to obtain 2-D visibilities, we Fourier transformed the synthetic images. We extracted 2 different baselines corresponding to the major and minor axis of the projected disk to provide the minimum and maximum visibility levels due to disk inclination. Interferometric visibilities obtained at 1 $\mu$m, 2.2 $\mu$m, 5 $\mu$m, 10 $\mu$m and 20 $\mu$m are presented respectively in Figs. 2, 3 and 4 for each model. The $x$-axis gives the projected baseline and the $y$-axis the normalized intensity. Values of visibilities corresponding to the disk major axis are given in Table 2 for projected baselines of 50 m and 100 m. We did not include visibilities at 0.5 $\mu$m since spectral energy distributions show that at this wavelength the star flux is dominant and therefore the disk is not resolved.

## 4. Discussion

### 4.1. Dynamic range issues

One of the main problems for detecting circumstellar disks by classical imaging techniques is usually the contrast in intensity between the star and the disk. How large is this problem for visibilities? The visibility of an unresolved object is a constant equal to one. An object is resolved if one can detect a slight decrease in the visibility curve. The diameter of a T Tauri star at 150 pc is 0.1 mas. The resolution of a 100-m baseline interferometer at $\lambda = 0.5$ $\mu$m is 1 mas. Therefore the star is always unresolved and the contribution to the visibility is a fixed level corresponding basically to the flux ratio of the star compared to the total flux of the T Tauri system. This ratio at different wavelengths is given by the SEDs where we can compute contributions from the star and from the disk. The position of line crossings gives the wavelength where the star and the disk fluxes are similar: at about $\lambda = 2$ $\mu$m for Model 1 and at about $\lambda = 1.4$ $\mu$m for Model 2. The stellar flux is completely embedded in Model 3. Therefore there is no dynamic range problem between the respective disk and star fluxes at wavelengths greater than 2 $\mu$m.

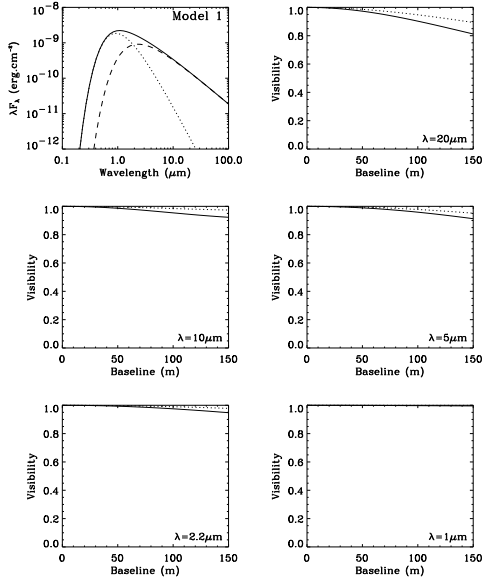

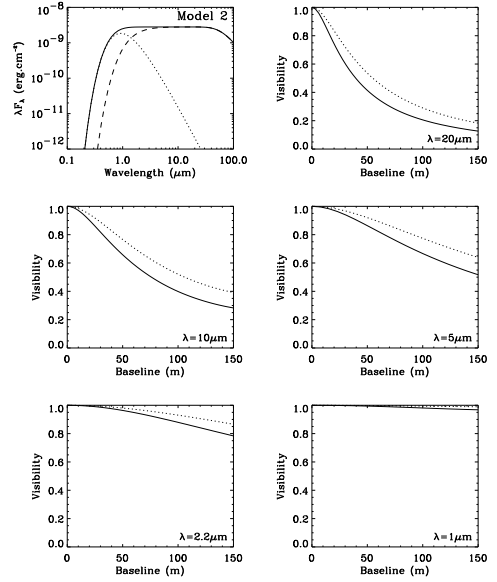

Fig. 2.— Interferometric visibilities of a classical T Tauri star (model 1). Top left panel: global spectral energy distribution in solid line, star contribution in dotted line and disk contribution in dashed line. Top right panel: visibility at 20 $\mu$m. Middle left panel: visibility at 10 $\mu$m. Middle right panel: visibility at 5 $\mu$m. Bottom left panel: visibility at 2.2 $\mu$m. Bottom right panel: visibility at 1 $\mu$m. Solid line corresponds to the major axis of the disk, and dotted line to the minor axis.

Fig. 3.— Interferometric visibilities of a T Tauri star with a flat infrared spectral energy distribution (Model 2). Top left panel: global spectral energy distribution in solid line, star contribution in dotted line and disk contribution in dashed line. Top right panel: visibility at 20 $\mu$m. Middle left panel: visibility at 10 $\mu$m. Middle right panel: visibility at 5 $\mu$m. Bottom left panel: visibility at 2.2 $\mu$m. Bottom right panel: visibility at 1 $\mu$m. Solid line corresponds to the major axis of the disk, and dotted line to the minor axis.

### 4.2. Resolution issues

In order to estimate at which wavelength one can start detecting circumstellar disk, one needs to know the accuracy of the visibility measurements. Thanks to careful calibration of the null spatial frequency, the precision of current interferometers is of order 1–5% for the visibility curves (cf. Pan et al. 1990 and Mozurkewitch et al. 1991 for *Mark III* stellar interferometer; Coudé du Foresto 1992 for *FLUOR* interferometer). From values reported in Table 2, T Tauri disk thermal emission is barely resolved for wavelengths shorter than 2.2 $\mu$m with a 100-m baseline in Model 1 and Model 2. However for Models 2 and 3, the detection of disk thermal emission is well within the range of optical interferometers at wavelengths longer than 1 $\mu$m.

Concerning the disk radial distribution of intensity, Figs. 2, 3 and 4 show that the longer the wavelength, the better T Tauri disks are resolved. The radial distribution of intensity at different wavelengths is displayed in Fig. 5 for Model 1, 2 and 3 in normalized units for the intensity and in milli-arcseconds for the distance to the star. The dotted lines represent the radial distribution of temperature in normalized units. The shapes of the solid curves can be separated into two asymptotic regimes. The first one corresponds to the Wien regime (inner radii) and the second one to the Rayleigh-Jeans regime (outer radii). The transition occurs at the maximum of the black-body emission when $\lambda T \approx 3000$ $\mu$m K. These transitions are marked by diamonds on the curves. The distribution of intensity is proportional to the distribution of temperature in the inner disk and decreases exponentially in the outer part of the disk. In the following we will call "disk size" the radius where the transition occurs.

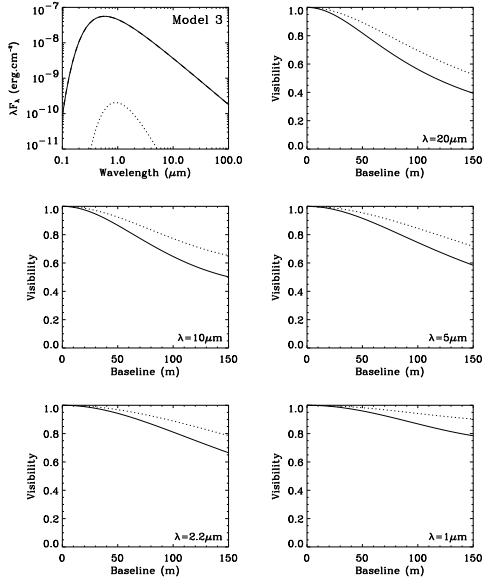

Fig. 4.— Interferometric visibilities of a FU Orionis star (Model 3). Top left panel: global spectral energy distribution in solid line, star contribution in dotted line and disk contribution in dashed line. Top right panel: visibility at 20 $\mu$m. Middle left panel: visibility at 10 $\mu$m. Middle right panel: visibility at 5 $\mu$m. Bottom left panel: visibility at 2.2 $\mu$m. Bottom right panel: visibility at 1 $\mu$m. Solid line corresponds to the major axis of the disk, and dotted line to the minor axis.

It is useful to make a connection between the disk size and the required interferometric resolution. A simple calculation shows that an interferometer of baseline $B$ can resolve at the 0.9 level a Gaussian disk of radius at half maximum given by the relation

$$R_d \geq 0.08\frac{\lambda}{B}. \qquad (1)$$

This criterion is not strictly valid for computed circumstellar disk models, but gives a useful first approximation of the needed resolution. Since the distribution of temperature is usually proportional to $R^{-q}$ with $q$ ranging between 0.5 and 0.75 (cf. Sect 2.), the exponential behavior of the distribution of intensity is smoother than a Gaussian ($e^{-r^q}$ compared to $e^{-r^2}$) and the resolution criterion is somewhat pessimistic for circumstellar disk models. Table 3 gives the corresponding size of the disk (defined by the size of transition zone) at 0.5 $\mu$m, 1 $\mu$m, 2.2 $\mu$m, 5 $\mu$m, 10 $\mu$m and 20 $\mu$m for the three models and the corresponding resolution of a 100-m baseline interferometer at the 0.9 level (cf. Eq. (1)). The resolution is marginal at 2.2 $\mu$m, 5 $\mu$m and 10 $\mu$m for Model 1, but the resolution of the interferometer is calculated with a pessimistic model and the level of detection of 0.9 is also a pessimistic estimate. A better estimate is found by assuming a realistic model and by computing the visibilities. This is done in Figs. 2, 3 and 4.

The wavelength behavior of the visibilities reflects the evolution of the resolution limit with the wavelength. In a classical disk, $T$ is proportional to $r^{-3/4}$, the limit of detection varies as $\lambda^{4/3}$. For a flare disk with $T \propto r^{-1/2}$, the limit of detection varies as $\lambda^2$. By combining the wavelength dependence of the disk size and of the interferometer resolution (which is proportional to the wavelength) one finds that the resolution of thermal emission from T Tauri disks varies as $\lambda^{1/3}$ for classical disks and as $\lambda$ for flare disks. This explains the difference of evolution

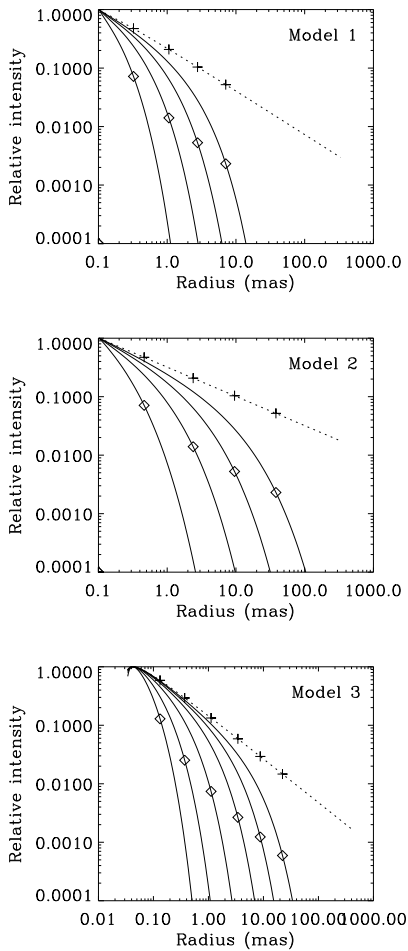

Fig. 5.— Radial distribution of intensity in the disk of a classical T Tauri star (Model 1, top), a T Tauri star with a flat infrared spectral energy distribution (Model 2, middle) and a FU Orionis star (Model 3, bottom). Solid lines from right to left: radial distributions of intensity at $\lambda = 20$ $\mu$m, $\lambda = 10$ $\mu$m, $\lambda = 5$ $\mu$m, $\lambda = 2.2$ $\mu$m, $\lambda = 1$ $\mu$m (Model 3 only) and $\lambda = 0.5$ $\mu$m (Model 3 only) normalized to unity at the inner radius of the disk. Diamond and plus marks correspond to the transition between the Wien and Rayleigh-Jeans regimes. Dotted line: radial distribution of temperature, normalized to unity at the inner radius of the disk.

with wavelength between Models 1 and 3, and Model 2.

The disk integrated thermal flux is proportional to the mass accretion rate (Bertout, Basri & Bouvier 1988). Therefore Model 3 is easier to resolve than Model 1, because the disk is hotter at a given radius. This result holds even if FU Orionis stars are generally located further ($d \geq 450$ pc) than the Taurus-Auriga cloud ($d \approx 150$ pc).

### 4.3. Sensitivity issues

The total system must be bright enough to obtain a sufficient signal-to-noise ratios in fringe measurements. If the signal is not strong enough, it becomes impossible to track the fringes which are moving because of the atmosphere disturbance. One generally defines a limiting magnitude or a limiting flux beyond which visibility cannot be measured.

The sensitivity limit for T Tauri disk detection is given by the total flux entering the lobe of the individual telescopes. We took as an example the case of the *Very Large Telescope Interferometer* in its *VISA (VLT Interferometric Sub-Array)* mode. VISA is composed of three 1.8-m auxiliary telescopes located on the *VLT* mountain with baselines ranging from 8 m to 200 m. Table 4 gives the sensitivity of the interferometer *VISA* (Beckers et al. 1989) and the expected fluxes from the three different models of T Tauri systems (calculated from simulations of the previous section). The *conservative* case corresponds to the state of the art in 1989 (detector noise, without adaptive optics,...) and with a signal-to-noise ratio equal to 50 for the fringe detection. The *long-term* case is more optimistic. It includes better performances of the detector, adaptive optics availability and a required signal-to-noise equal to 10 for fringe tracking. The first case represents the expected initial performance of VISA and the second case the expected fully developed performance.

Table 3: Disk sizes vs. resolution of a $B = 100$ m interferometer at the 0.9 level.

| Wavelength | Model 1 | | Model 2 | | Model 3 | | 0.08 $B/\lambda$ |
| --- | --- | --- | --- | --- | --- | --- | --- |
| | AU | mas | AU | mas | AU | mas | mas |
| $\lambda = 0.5$ $\mu$m | – | – | – | – | 0.06 | 0.1 | 0.08 |
| $\lambda = 1$ $\mu$m | – | – | – | – | 0.15 | 0.4 | 0.16 |
| $\lambda = 2.2$ $\mu$m | 0.05 | 0.3 | 0.07 | 0.5 | 0.5 | 1 | 0.4 |
| $\lambda = 5$ $\mu$m | 0.15 | 1 | 0.4 | 3 | 1.5 | 3 | 0.8 |
| $\lambda = 10$ $\mu$m | 0.4 | 3 | 1.5 | 10 | 4 | 9 | 1.6 |
| $\lambda = 20$ $\mu$m | 1 | 7 | 6 | 40 | 10 | 20 | 3.3 |

Table 4: T Tauri accretion disks predicted fluxes compared to VISA expected sensitivity

| Wavelength | T Tauri disk fluxes | | | VISA Sensitivity | |
| --- | --- | --- | --- | --- | --- |
| | Model 1 | Model 2 | Model 3 | *conservative case* | *long-term case* |
| 0.5 $\mu$m | 0.13 Jy | 0.13 Jy | 9.1 Jy | 4.2 Jy | 21 mJy |
| 1 $\mu$m | 0.73 Jy | 0.83 Jy | 15 Jy | 1.4 Jy | 21 mJy |
| 2.2 $\mu$m | 1.2 Jy | 2.1 Jy | 16 Jy | 0.4 Jy | 8 mJy |
| 5 $\mu$m | 1.2 Jy | 4.7 Jy | 14 Jy | 0.5 Jy | 0.1 Jy |
| 10 $\mu$m | 1.1 Jy | 9.3 Jy | 12 Jy | 1.6 Jy | 0.3 Jy |
| 20 $\mu$m | 1.0 Jy | 19 Jy | 9.9 Jy | 5.6 Jy | 1.1 Jy |

The comparison shows that T Tauri disks are detectable with VISA expected performances. The detection will only be possible beyond 2.2 $\mu$m for the weakest system (Model 1) at the beginning, but in the long term will not be a problem. For brighter systems like flat infrared spectrum T Tauri stars (Model 2) or FU Orionis stars (Model 3), VISA will sufficiently sensitive even in the first stage of development.

### 4.4. Conclusion

We have limited our study to thermal detection of circumstellar disks around T Tauri stars. We have shown that one can detect and resolve these disks at wavelengths longer than 2.2 $\mu$m, for example with the *VISA* interferometer. No better spatial resolution will be achieved in the near-infrared with other instruments than long baseline interferometers, except perhaps a lunar 100-m telescope. Optical long baseline interferometry is therefore the only technique which permits detection of thermal emission of T Tauri disks in the optical wavelengths.

### REFERENCES

Adams, F.C., Emerson, J.P., Fuller, G.A. 1990, ApJ 357, 606

Adams, F.C., Lada, C.J., Shu, F.H. 1988, ApJ 326, 865

Appenzeller, I., Jankovics, I., Oestreicher, R. 1984, A&A 141, 108

Basri, G., Bertout, C. 1989, ApJ 341, 340

Beckers, J., Braun, R., di Benedetto, G.P., Foy, R., Genzel, R., Koechlin, L., Merkle, F., Weigelt, G. 1989, ESO/VLT Report 59b, 10

Beckwith, S.V.W., Sargent, A.I., Chini, R.S., Güsten, R. 1990, AJ, 99, 924

Bertout, C., Basri, G., Bouvier, J. 1988, ApJ 330, 350

Bertout, C., Bouvier, J. 1988, in proceedings of NOAO/ESO conference on "High Resolution Imaging by Interferometry", ed. Merkle, F., 69

Bester, M., Danchi, W.C., Degiacomi, C.G., Bratt, P.R., 1994, SPIE 2200, 29

Bouvier, J., Bertout, C. 1992, A&A 263, 113

Bouvier, J., Malbet, F., Monin, J.-L. 1992, Ap&SS 212, 159

Cabrit, S., Edwards, S., Strom, S.E., Strom, K.M. 1990, ApJ 354, 687

Coudé du Foresto, V., Ridgway, S., and Mariotti J.-M. 1992 in Proceedings of the ESA Colloquium on "Targets for Space-based Interferometry", Beaulieu, ed. Volonté, S., ESA SP-354, 219

Edwards, S., Cabrit, S., Strom, S.E., Heyer, I., Strom, K.M., Anderson, E. 1987, ApJ 321, 473

Ferreira, J., Pelletier, G. 1994, A&A, in press

Lynden-Bell, D., Pringle, J.E. 1974, MNRAS 168, 603

Malbet, F., Bouvier, J., Monin, J.-L. 1992, in proc. of ESA Colloquium on "Targets for Space-Based Interferometry", ed. Volonté, S. (ESA SP 354), 111

Monin, J.-L., Bouvier, J., Malbet, F. 1993, in proc. of IAU Symposium 158 on "Very High Angular Resolution Imaging", eds. Robertson, J.G. & Tango, W.J., 387

Mozurkewich, D., Johnson, K.J., Simon, R.S., Bowers, P.F., Gaume, R. et al. 1991, AJ, 101, 2207

Pan, X.P., Shao, M., Colavita, M.M., Mozurkewich, D., Simon, R.S., Johnson, K.J. 1990, ApJ 356, 641

Roddier, F. 1979, J.Opt. (Paris) 10, 299

Rydgren, A.E., Zak, D.S. 1987, PASP 99, 191

Shu, F.H., Najita, J., Ostriker, E., Wilkin, F., Ruden, S.P., Lizano, S. 1994, ApJ 429, 781

Townes, C.H., 1984, JA&A, 4, 111